\documentclass[%
 aip,
 amsmath,amssymb,
reprint, onecolumn 
]{revtex4-1}

\usepackage{graphicx}
\usepackage{dcolumn}
\usepackage{bm}

\usepackage[utf8]{inputenc}
\usepackage[T1]{fontenc}
\usepackage{mathptmx}
\usepackage{etoolbox}
\usepackage{dcolumn}
\usepackage[colorinlistoftodos]{todonotes}
\makeatletter
\def\@email#1#2{%
 \endgroup
 \patchcmd{\titleblock@produce}
  {\frontmatter@RRAPformat}
  {\frontmatter@RRAPformat{\produce@RRAP{*#1\href{mailto:#2}{#2}}}\frontmatter@RRAPformat}
  {}{}
}%
\makeatother

\usepackage{xcolor}
\usepackage{color,soul}

\begin{document}

\preprint{AIP/123-QED}

\title{Surfactant-Driven Dynamic Changes in Rheology of Activated Carbon Slurry Electrodes}
\author{Mohan Das}
 \altaffiliation[Also at ]{Faculty of Physics, University of Vienna, Boltzmanngasse 5, A-1090, Vienna, Austria}
\author{KangJin Lee}%

\author{Christopher L. Wirth}
 \email{wirth@case.edu.}
\affiliation{
Department of Chemical and Biomolecular Engineering, Case Western Reserve University, Cleveland, Ohio 44106.
}%

\date{\today}

\begin{abstract}
\textit{Slurry electrodes are an effective means to improve flow battery performance by reducing electrode fouling and increasing the active surface area necessary for electrochemical reactions. It is critical to understand how the slurry formulation impacts its rheological profile and ultimately battery performance. We study the linear and nonlinear rheology of activated carbon (AC) based slurry electrode used in an all-iron flow battery in the presence of a nonionic surfactant (Triton X-100). Our results show the slurry mimics a colloidal gel with elasticity remaining constant despite increasing surfactant concentration until $\alpha$(=C$_{surf.}$/C$_{AC}$) < 0.65. However, at $\alpha \ge$ 0.65, the slurry abruptly transitions to a fluid with no measurable yield stress. This critical surfactant concentration at which the rheological profile undergoes a dynamic change matches the concentration found previously for gel collapse of this system. Moreover, this transition is accompanied by a complete loss of electrical conductivity. These data show that site specific adsorption of surfactant molecules often used in slurry formulation have a significant and dramatic impact on flow behavior, which should be considered when formulating a slurry electrode.}
\end{abstract}

\maketitle

\section{\label{sec:1}Introduction}

Finding long-term solutions for environmentally sustainable production and storage of energy has become a global priority. Flow battery technology is considered crucial for energy storage applications needed for the transition to renewable energy production, especially those that require load leveling and peak shaving.\cite{wang2013recent,park2016material} Flow batteries have several advantages such as modular design, moderate cost of operation and maintenance, and the decoupling of capacity and power.\cite{de2006redox} Energy storage capacity in a flow battery is determined by the volume and type of the electrolytes stored in separate reservoirs whereas power rating is determined by the efficiency and surface area of the electrochemical cell.\cite{wang2013recent} Redox flow batteries (RFBs) store charge via an electrochemical redox reaction at their electrodes, such as in the case of the all-vanadium flow battery.\cite{zhang2019progress} More recently, the use of flowable electrodes in the form of conductive particles suspended in the electrolyte has been shown to improve battery performance.\cite{boota2015graphene,lohaus2019charge,percin2020resistances, LEE2023712} Carbon based particles are commonly used as the conductive medium due to their ability to conduct electrons by forming a continuous network of particles as well as their ability to conduct ions from the electrolyte.\cite{campos2013investigation,presser2012electrochemical,mourshed2021carbon,tam2023electrochemical} All-iron flow batteries are considered a cost-effective solution compared to other common types of flow batteries due to the abundance and benign nature of iron.\cite{hruska1981investigation,petek2015slurry}

\par All-iron flow batteries often use carbon particles as the conductive medium.\cite{petek2015slurry,tam2023electrochemical} Notably, activated carbon (AC) particles have comparatively higher porosity and surface area compared to other types of carbon particles.\cite{burchell1999carbon} These features are helpful to energy storage applications because carbon acts as an electron conducting medium as well as the site of Faradaic redox reactions.\cite{mourshed2021carbon} Moreover, carbon has a wide operational potential range, low cost, and good chemical stability when compared to metal particles.\cite{chakrabarti2014application} However, the use of carbon particles in aqueous electrolytes presents a unique set of challenges. In general, it is difficult to disperse carbon particles due to their surface hydrophobicity\cite{ridaoui2006effect} and suspension structure important to electronic conductivity will be impacted by shearing flow during flow battery operation. Strategies to increase suspension conductivity, such as increasing particle concentration, can be problematic because of clogging. \cite{presser2012electrochemical,mourshed2021carbon} Although AC particles are typically more hydrophilic than regular carbon black particles, AC particles tend to aggregate due to van der Waals forces at low pH (< 2) as their surface charge is neutralized.\cite{park2016material} At low particle volume fraction, when the suspension is at rest (or when the battery is not operational), particle aggregation can lead to sedimentation and clogging of flow channels.\cite{presser2012electrochemical,LEE2023712} Moreover, in the case of carbon black particles, studies have reported that when the particle volume fraction is increased above a percolation threshold the particles can form three dimensional continuous networks.\cite{helal2016simultaneous,richards2017clustering} Networking of particles is thought to improve electronic conductivity, with the percolation threshold often referred to as electrical percolation.\cite{youssry2013non,helal2016simultaneous} Particle networks can also become strong enough to bear mechanical stresses. This threshold is known as mechanical percolation or gelation.\cite{lu2008gelation} In the context of colloidal stability, gelation is a useful tool to arrest or slow-down sedimentation of solid particles and is regularly used in the design of consumer products, food, and cosmetics.\cite{poon1999delayed,manley2005gravitational,kirilov2014aqueous,chelazzi2018microemulsions} Unfortunately, mechanically percolated suspensions often have complex rheological profiles that would present challenges in the design of slurry flow batteries that usually have complex flow geometries. For instance, gel forming suspensions, such as carbon black particles, are extremely sensitive to flow rates in a way that can result in shear dependent rheological behavior.\cite{osuji2008shear,ovarlez2013rheopexy,narayanan2017mechanical,hipp2021direct,dages2022interpenetration,richards2023review} Reliable operation of a slurry flow battery requires that the slurry electrode is stable and does not clog the flow geometry. Moreover, it is important that the slurry electrode does not sediment or be highly viscous leading to higher pumping cost affecting battery efficiency.

\par Although limited, there are examples of efforts to control the stability and rheological profiles of slurry electrodes in flow batteries. For example, studies have reported the use of surfactants in slurry electrodes to modify their flow behavior, with the main aim to improve particle dispersion reducing the slurry viscosity.\cite{madec2015surfactant,lee2016use} Studies have found that nonionic surfactants have higher affinity to adsorb to carbon based particles in an aqueous medium and do not interfere with the redox reaction.\cite{sis2009effect} Adsorbed surfactants improve particle dispersion by reducing interparticle aggregation whereby surfactant monomers form a steric layer on the particle surface. Reduction in aggregate hydrodynamic size and interparticle repulsion would result in reduced slurry viscosity.\cite{hipp2021direct} Limited studies exist for highly porous particles such as AC particles, especially within the context of how surfactant adsorbs to the very high surface area particle and results in changes of rheology. Further, there is limited work on determining how surfactant adsorption for these systems impacts electrical conductivity and performance of the flow battery.

\par In this study, we measured the impact of nonionic surfactant (Triton X-100) on the stability and rheological response of slurry electrode composed of AC particles dispersed in an aqueous electrolyte. We found the addition of surfactant had limited impact on slurry viscoelasticity below a critical surface coverage of surfactant, with the concentration corresponding to that previously found to induce catastrophic collapse.\cite{LEE2023712} Beyond this critical coverage, the slurry viscoelasticity changed sharply, rather than monotonically. Further, we found the additional electrical conductivity provided by the AC particles to be eliminated above this critical coverage. This behavior of slurry viscoelasticity with increasing surfactant concentration indicates a unique surfactant adsorption mechanism and its effect on interparticle attraction affecting rheology and conductivity.


\section{\label{sec:2}Materials and methods}
\subsection{\label{sec:2.1}Preparation of slurry electrode}
The formulation of the slurry electrode used herein matches that used previously for an all-iron flow battery.\cite{petek2015slurry} Briefly, samples were prepared by adding YP-50 activated carbon black particles (Kuraray Co. Ltd.) in 1 M H$_2$SO$ _4$ containing 0.55 M Fe$_2$SO$_4$. The particles were dispersed by stirring using magnetic pellets overnight in 20 ml glass vials. Two sets of samples were prepared. In the first set, carbon black was varied from 6 - 16\% w/w. In the second set, carbon black concentration was fixed at 14 \% w/w and a nonionic surfactant, Triton X-100 (Sigma Aldrich, M.W. = 625 g/mol) was added to 1 M H$_2$SO$_4$ containing 0.55 M Fe$_2$SO$_4$. The ratio of concentration of surfactant to that of carbon black denoted here as $\alpha$ (= c$_{surf}$/c$_{AC}$) varied from 0 - 0.7. Note here that very low pH of the acid solution as well as the high ionic strength due to Fe$_2$SO$_4$ addition will lead to the screening of the electrostatic double layer repulsion between AC particles. We expect this screening to promote interpartcle aggregation.

\subsection{\label{sec:2.2}Adsorption isotherm measurements}
Samples were prepared at a fixed particle concentration of 14\% w/w to measure the adsorption of the nonionic surfactant on carbon black particles. Here, $\alpha$ was varied from 0.4-1.4. After mixing the samples overnight, the slurry was centrifuged (Sorvall ST 8, ThermoFisher) at 5000$\times$g for 10 minutes to allow AC particle sedimentation. The collected supernatant was filtered (Fisherbrand - PTFE, 450nm pore size) to remove any remaining AC particles. The concentration of the free surfactant was measured using a UV-vis spectrophotometer (Cary 3500, Agilent) in the wavelength range of 200-500 nm. For all the measurements, samples were diluted 100$\times$ with 1 M H$_2$SO$_4$ to avoid saturation limit of the equipment. Therefore, 1M H$_2$SO$_4$ with 5.5 mM FeSO$_4$ was used as the baseline solvent to remove effect of FeSO$_4$ in the system.

\subsection{\label{sec:2.3}Conductivity Measurements}
Samples were prepared in 250 ml glass bottles to measure the overall conductivity of the slurry. Here the AC particle concentration was fixed at 10\% w/w at $\alpha$ = 0 and 0.7. The flow rate of the slurry was fixed at 170 ml min$^{-1}$ and was pumped with a table top peristaltic pump (Masterflex L/S Peristaltic Pump, Cole Parmer) connected by 0.25 inch ID silicone tubing through an unseparated redox flow cell. The slurry reservoir was constantly stirred with a magnetic stir bar during the measurement to avoid aggregation and sedimentation of the AC particles. A potentiostat (Reference 600, Gamry Instruments) was used to carry out cyclic voltammery (CV) measurements which were conducted over a potential range of $\pm$ 0.1V vs open circuit potential with scan rates ranging from 1 mV/s - 1 V/s. The slope of the CV curve was used too calculate the conductivity based on the geometry of the cell (0.08 cm channel thickness with surface area of 3 cm$^2$). 

\subsection{\label{sec:2.4}Rheological protocol}

Rheological measurements were carried out using a stress-controlled rheometer (Haake Mars 3, Thermoscientific). The sample was loaded between a serrated parallel plate geometry. The plates have a diameter of 35 mm and a gap of 400 $\mu$m. Sample temperature was maintained constant at 25 $\pm$ 0.1 $^{\circ}$C using a peltier control element. The gap between the plates was fixed at 400 $\mu$m after measuring sample viscosity at different gaps and determining that confinement effects were eliminated at this gap. Note here that the shear rates applied correspond to the shear rate at the outer edge of the plate-plate geometry. The whole geometry was covered with a solvent trap and a pool of water was maintained around the bottom plate using vacuum grease as a barrier. The pool of water ensures a humid environment within the trap. This setup prevented sample evaporation for the duration of the experiment.
\par After loading, the sample was presheared at a steady shear rate of 1000 s$^{-1}$ for 300 s to remove the effect of shear history during sample loading, a process known as rejuvenation. 
The rejuvenation step was followed by shear flow cessation for 2000 s. During this time a small amplitude oscillatory shear (SAOS) was applied at a strain amplitude $\gamma$ = 0.2\% and angular frequency $\omega$ = 10 rad.s$^{-1}$ to capture the time evolution of sample elastic (G') and viscous modulus (G'') known as \textit{aging}. Our measurements have shown that 2000 s were enough for the shear moduli to reach steady state. This was followed by a dynamic frequency sweep (DFS) measurement carried out at $\gamma$ = 0.2\% and $\omega$ = 1-100 rad.s$^{-1}$. Aging and DFS measurements were followed by dynamic strain amplitude sweep (DSS) measurement carried out at $\gamma$ = 0.1 - 1000\% and $\omega$ = 10 rad.s$^{-1}$ to investigate the yielding (solid to liquid-like transition) of the slurry. All the previous steps were repeated three times to evaluate repeatability.
\par The steady shear response of the slurry was determined by performing an applied constant shear rate (shear startup) and shear stress (creep) measurements. The flow curve was determined by performing an upward and downward shear rate sweep, $\dot{\gamma}$ = 0.1-1000 s$^{-1}$ with the shear stress $\sigma$ recorded at 15 s after the applied shear rate. The average values of stress are reported for downward shear rate sweeps repeated three times. We only used the $\sigma$ values obtained during downward shear rate sweep to plot the flow curve. Shear startup measurement was performed by first rejuvenating the sample and then stopping the flow for 600 s followed by applying a constant shear rate for 1000 s and measuring the resultant stress. Creep measurements were performed by first rejuvenating the sample and then stopping the flow for 600 s followed by applying a constant shear stress for 1000 s and measuring the resultant shear rate. The parameters influencing the slurry flow behavior were determined using modified Herschel-Bulkley (HB) model commonly used for similar systems:\cite{herschel1926, hipp2021direct}
\begin{equation}
\label{Eq:Eq1}
    \sigma(\dot{\gamma}) = \sigma_{y1} \left( 1+\left(\dfrac{\dot{\gamma}}{\dot{\gamma_c}}\right)^n \right) \quad \textrm{for} \quad |\sigma| > \sigma_y,
\end{equation}
where $\sigma_{y1}$ is the dynamic yield stress, $\dot{\gamma_c}$ is the critical shear rate and \textit{n} is the power-law index. Note here that in samples with added surfactant, the solvent viscosity $\eta_s$ was measured after particle sedimentation using centrifugation at 5000$\times$\textit{g}.

\section{\label{sec:3}Results}
\subsection{\label{sec:3.1} Particle morphology}
\begin{figure*}[ht]
\includegraphics[width=0.6\linewidth]{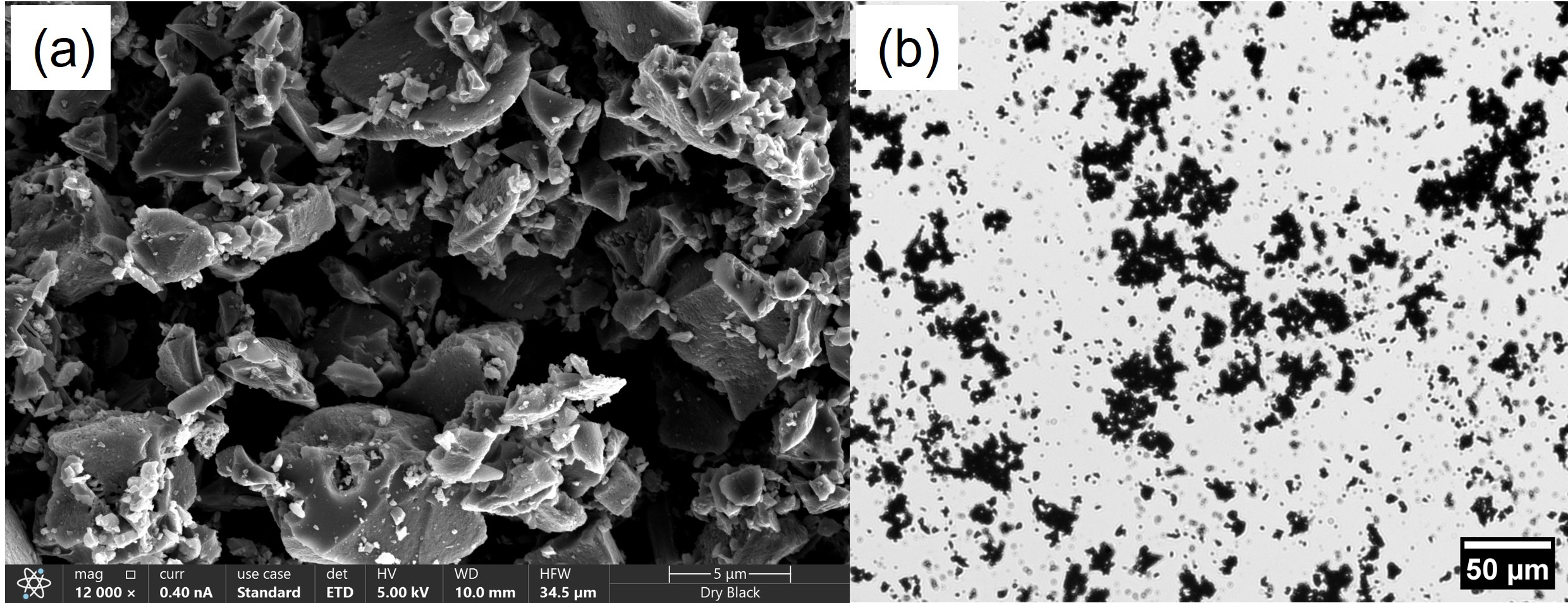}
\caption{\label{fig:fig1} (a) Scanning electron microscope image of dry YP-50 AC particles, (b) Brightfield microscopy image of 1 \% w/w of AC particles dispersed in 1 M H$_2$SO$_4$ containing 0.55 M Fe$_2$SO$_4$.}
\end{figure*}

YP-50 activated carbon (AC) is prepared by hydrothermal activation of coconut-shell resulting in very high surface area, high electrical conductivity, and chemical stability.\cite{jain2013activated} The very high surface area available for the electrochemical reaction is a result of the large porosity in this type of AC and results in complex particle morphology as seen in Fig. \ref{fig:fig1} (a). However, the synthesis process also renders high polydispersity in particle size with a range of 5-20 $\mu$m.\cite{boota2014activated} AC particles typically disperse better in aqueous medium compared to other carbonaceous particles due their chemical activation. Chemical activation such as alkaline activation commonly used for AC relevant to electrochemical applications renders their surface hydrophilic due to the presence of oxygen containing carboxylic, anhydride and lactonic groups.\cite{gao2020insight} This also makes their surface charged when dispersed in water at neutral pH.\cite{brennan2002adsorption} However, dispersing AC in very low pH and high ionic strength media such as used herein would neutralize their negative surface charge\cite{LEE2023712} causing interparticle aggregation through van der Waals attraction and results in particle clusters as shown in Fig. \ref{fig:fig1} (b). It is well known that in attractive systems, particle suspensions can form a metastable state known as gel when particle concentration is increased above a certain threshold known as the critical gel point.\cite{eberle2011dynamical,helgeson2014homogeneous} One would expect gelation to significantly affect the flow behavior of a slurry electrode and by extension impact the overall battery performance. Thus, we're interested in improving our understanding of the rheological response of slurry to changes in formulation. Next, we explore the influence of AC concentration on slurry rheology.

\subsection{\label{sec:3.2} Effect of particle concentration on slurry viscoelasticity}

\begin{figure*}[ht]
\includegraphics[width=0.9\linewidth]{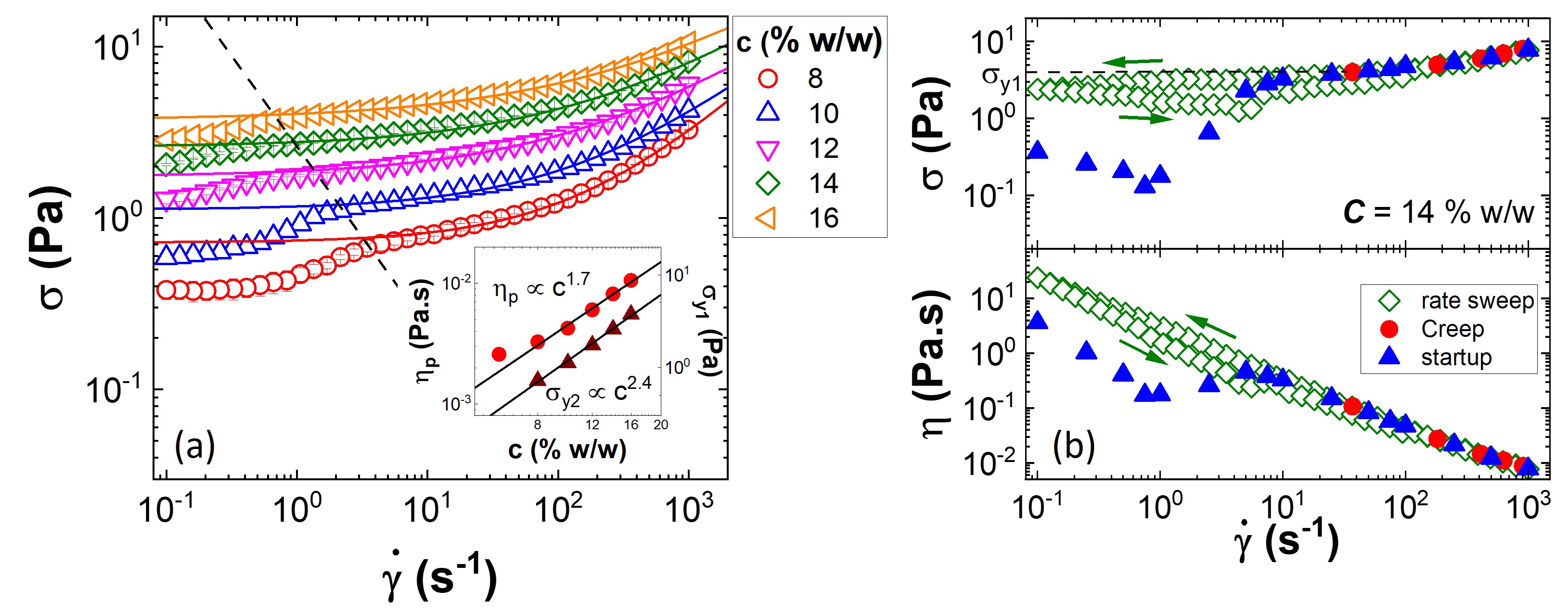}
\caption{\label{fig:fig2} (a) Flow curve plotted as $\sigma$ vs $\dot{\gamma}$ for slurry samples at different particle concentration, \textit{C}. Solid line is the Herschel-Bulkley fit. Dotted line represents the boundary of $\dot{\gamma}$ below which we observe wall-slip (inset: Apparent viscosity $\eta_p$ measured at 1000 s$^{-1}$ and $\sigma_{y1}$ from Herschel-Bulkley fit for different \textit{C}. Black lines are best power-law fits with a slope of 1.7 and 2.4 for $\eta_p$ and $\sigma_{y1}$ respectively.) (b) intrinsic flowcurve hysteresis measured for 14\% w/w suspension showing counter-clockwise loops in shear stress values (recorded at 15s after applied shear rate). Filled symbols indicate steady state values of $\sigma$ and $\dot{\gamma}$ obtained from shear startup and creep  measurements respectively.}
\end{figure*}

Fig. \ref{fig:fig2} (a) shows the flow curve plotted for a series of particle concentrations considered herein. The shear stress $\sigma$ plotted against shear rate $\dot{\gamma}$ exhibits behavior generally observed with colloidal gels. Namely, we found the shear stress increased with a power law slope at high shear rates and exhibited a clear yield stress plateau at low shear rates. The stress response data was fit reasonably well using the modified HB model shown in Eq. \ref{Eq:Eq1} and the parameters determined are listed in Table \ref{tab:table1}. Note here that shear rates below which the shear stress drops sharply were not considered for the fitting. HB yield stress $\sigma_{y1}$ values scale with a power of 2.4 when plotted against particle concentration (inset: Fig. \ref{fig:fig2} (a)) which is slightly lower than the values reported in literature for carbon black particle gels.\cite{grenard2014timescales,n2020yielding} Increased particle concentration resulted in a higher yield stress from the larger number of networks formed by particle flocs in this attractive system. We also measured the slurry viscosity $\eta_p$ at the highest applied shear rate of 1000 s$^{-1}$ and see that it scales with a power-law of 1.7 with particle concentration. 
\par We calculated P\'{e}clet number Pe = $\dfrac{\sigma a^3}{kT}$ using the mean particle size and shear stress measured from the flowcurve and found that Pe $\approx$ 10$^4$-10$^5$ indicating that hydrodynamic forces dominate over Brownian forces in the entire range of shear rates applied.\cite{osuji2008shear} In other words, this means the AC particles cannot reorganize within a cluster through self-diffusion even at the lowest shear rates applied here. 
\par Note another important feature is where we observed a drop in shear stress at low shear rates (Fig. \ref{fig:fig2} (a)). This has been observed in other attractive systems including carbon black suspensions.\cite{n2020yielding,hipp2019structure} We initially  attributed this to the wall-slip or shear banding phenomenon very commonly observed in attractive suspensions.\cite{cloitre2017review,abbasi2022apparent} Moreover, this critical shear rate reduces with increasing particle concentration (as indicated by dotted line in Fig. \ref{fig:fig2} (a). However, on closer observation we captured a hysteresis in the flow curve at low shear rates while applying forward and reverse shear rate sweep (Fig. \ref{fig:fig2} (b), open symbol). Here the stress had higher values in the reverse shear rate sweep compared to forward shear rate sweep with the hysteresis loop having a counter-clockwise direction. Additionally, when a shear startup measurement was performed under constant applied shear rate, the shear stress decayed over time (Fig. S1, ESI) in the shear rate regime which we previously attributed to wall-slip in Fig. \ref{fig:fig2} (a). This behavior has been reported for carbon black suspensions and is thought to be a rheological signature of rheopexy or antithixotropic dynamics.\cite{ovarlez2013rheopexy,wang2022new} \par Similarly during creep measurement, the suspension does not flow ($\dot{\gamma}\to 0$) below a critical shear stress (Fig. S2, ESI) which coincides with the HB yield stress. When the applied shear stress is above the yield stress, it results in a very large increase in shear rates indicating a rapid fluidization of the suspension. These rheological data indicate a sharp transition of AC particle suspensions from the weak-flow limit to the strong-flow limit over a very narrow range of shear stresses. This suggests that our slurry flow behavior is similar to that of carbon black gels such that at high shear rates or stresses the AC particle clusters break down into smaller aggregates, whereas at low shear rates or stresses the AC tend to form shear-driven agglomerates.\cite{hipp2021direct,wang2022new} Moreover, this structural effect is more prominent at lower particle concentrations, which is more prone to microstructural heterogeneity.

\begin{table}[ht]
\caption{\label{tab:table1}Parameters used for Herschel-Bulkley model fits to the suspension flow curve }
\begin{ruledtabular}
\begin{tabular}{ccdd}
$C (\% w/w)$ & $\sigma$ (Pa) & \mbox{$\dot{\gamma_c}$ (s$^{-1}$)} & \mbox{n} \\
\hline
8 & 0.72 $\pm$ 0.008 & 166 & 0.7 \\
10 & 1.12 $\pm$ 0.01 & 185 & 0.6 \\
12 & 1.75 $\pm$ 0.015 & 187 & 0.5 \\
14 & 2.6 $\pm$ 0.02 & 227 & 0.5 \\
16 & 3.75 $\pm$ 0.03 & 281 & 0.45 \\
\end{tabular}
\end{ruledtabular}
\end{table}

\begin{figure*}[ht]
\includegraphics[width=0.9\linewidth]{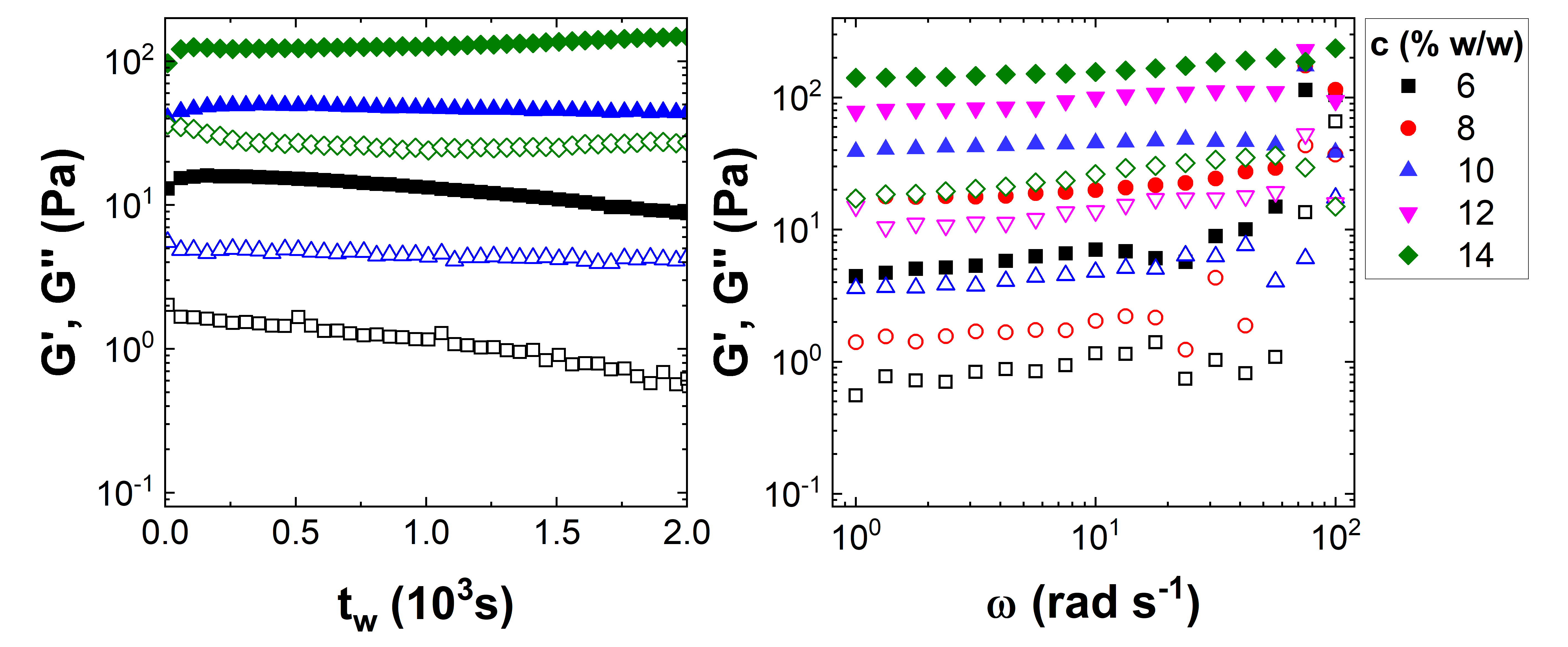}
\caption{\label{fig:fig3} (a) Time-evolution of elastic modulus G' (closed) and viscous modulus G'' (open) at different particle concentration \textit{C} measured at $\omega$ = 10 rad.s$^{-1}$ and t$_w$ = 2000 s, (b) dynamic frequency sweep (DFS) measurement for different particle concentration \textit{C}.}
\end{figure*}

Considering the narrow geometry encountered within a flow battery, our calculations show that the shear rate within the cell can exceed 1000 s$^{-1}$ under typical operating conditions. We have seen that at such high shear rates, the AC particle slurry is rapidly fluidised. Interestingly, colloidal gels when fluidised after subjecting to high shear rates tend to restructure on shear cessation and recover their solid-like character.\cite{koumakis2015tuning} We measured  the time-evolution of viscoelasticity (or ageing) as shown in Fig.\ref{fig:fig3} (a) after subjecting the sample to a high shear rejuvenaton at 1000 s$^{-1}$ for 300 s. We observed the onset of gelation characterised by the cross over of storage modulus G' over loss modulus G'' at waiting time t$_w$ < 10 s. This indicated a rapid percolation of particle aggregates after cessation of shear and as seen here is independent of particle concentration. This behavior is observed in gels with van der Waals attraction force as the primary driver of interparticle aggregation.\cite{das2022shear} In most cases (except at 6\% w/w) elastic modulus increased over time albeit not sharply, before reaching a quasi-steady state indicating that the particle microstructure reached its final spatial distribution. This is in contrast with aging behavior displayed by carbon black particles dispersed in an aqueous medium where the elastic modulus increased by a factor of over 10 times within an hour and did not reach a steady state.\cite{n2020yielding} However, for 6\% w/w sample here, we observed a drop in shear moduli that can be attributed to particle sedimentation. In attractive systems, the gravitational stresses could be greater than network elasticity at dilute particle concentrations leading to sedimentation within the experimental time window.
 
\par We measured the response of shear moduli to frequency by performing a dynamic frequency sweep (DFS) measurement over two decades of frequency after allowing the sample to age for 2000 s. As can be seen from Fig. \ref{fig:fig3} (b), both elastic and viscous moduli show a very weak increase with frequency $\omega$ with a platueau of G' at lower frequency. In all the samples measured here, they exhibited a clear solid-like response (G' > G'') with the absence of terminal relaxation for G' as is the case for colloidal gels.\cite{koumakis2011two,laurati2011nonlinear} DFS measurement also shows that at the lowest particle concentration studied here (6\% w/w) we form a stable gel. 

\par
\begin{figure*}[ht]
\includegraphics[width=0.9\linewidth]{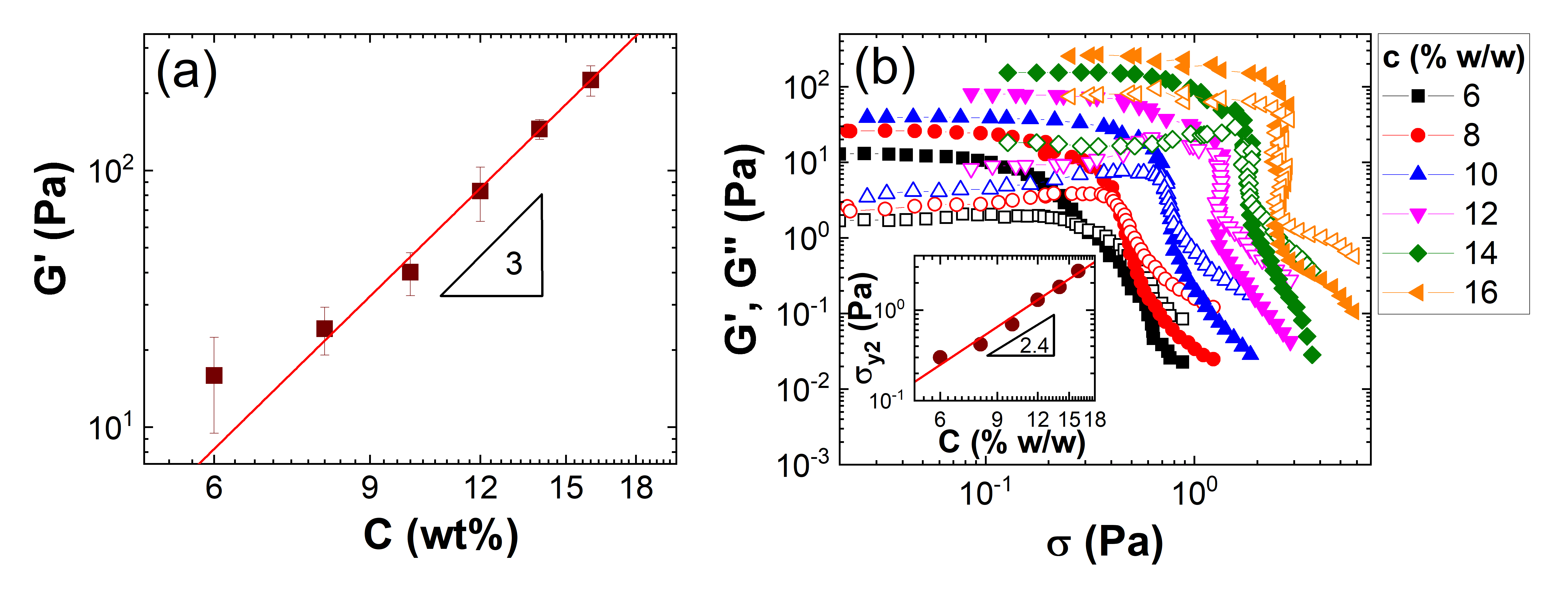}
\caption{\label{fig:fig4} (a) Elastic modulus G' plotted against particle concentration \textit{C}, solid line is the power law fit for G' $\sim$ C$^3$ excluding 6\% w/w sample due to sedimentation, (b) evolution of elastic and viscous moduli during DSS measurement plotted against measured stress $\sigma$ at different particle concentration \textit{C} (inset: yield stress measured from cross over of G' and G'' from DSS, $\sigma_{Y2}$ plotted against particle concentration \textit{C} with a best power-law fit with exponent 2.4).}
\end{figure*}

 Previous studies on colloidal gels have found a now well-established power-law scaling relationship between the elastic modulus and particle concentration, G' $\sim$ \textit{C}$^{\beta}$ with values ranging from 2-4.5 depending on interparticle potential.\cite{trappe2001jamming,buscall1988mills,dages2022interpenetration} In our case, we found that for AC slurry that restructures after rejuvenation, G'$\sim$ \textit{C}$^{3}$ (Fig. \ref{fig:fig4} (a)) and is similar to that reported for carbon black gels.\cite{trappe2000scaling,grenard2014timescales} Additionally, taking into account the preshear stress $\sigma_{p}$ during rejuvenation, this relationship can be further interpreted as G'$\sim$ \textit{C} $\sigma_{p}^{d_{f}/(3-d_{f})}$ where d$_f$ is the cluster fractal dimension.\cite{osuji2008shear} This relationship is in turn derived from R$_c \sim \sigma_{p}^{1/(3-d_{f})}$ where R$_c$ is the preshear stress dependent fractal cluster radius and G' $\sim$ \textit{CU}/R$_c^{d_f}$ where U is the interparticle potential.\cite{osuji2008shear,dages2022interpenetration} Considering that preshear viscosity $\eta_p$ = $\sigma_{p}/\dot{\gamma_p}$, where $\dot{\gamma_p}$ = 1000 s$^{-1}$ is the rejuvenation shear rate, and G' $\sim$ \textit{C}$^{\beta}$, we get $\eta_p \sim$ \textit{C}$^{(\alpha-1)(3-d_{f})/d_{f}}$. From inset: Fig. \ref{fig:fig2} (a), we see that $\eta_p \sim$ \textit{C}$^{1.7}$. Taking $\beta$ = 3, we obtain d$_f$ = 1.7 which indicates strong links between particles in the gel backbone as can be expected for particle gels formed by van der Waals attraction.\cite{dages2022interpenetration} The above results clearly indicate that after cessation of rejuvenation, the AC slurry forms a fractal like gel structure with strong interparticle links similar to carbon black gels and follows very well the scaling of G' with \textit{C} and $\sigma_p$.

\par To further investigate the solid-liquid transition (known as yielding) of gel formed by AC particles, we performed an oscillatory shear test with increasing stress amplitude and a frequency of 10 rad.s$^{-1}$. Applying stress amplitude is beneficial here to prevent feed back fluctuations commonly observed in stress-controlled rheometers when performing strain amplitude sweep. As seen in Fig. \ref{fig:fig4} (b), both elastic and viscous moduli remain unchanged at low stress amplitudes. However, as the stress amplitude is increased we see that both the elastic and viscous modulus drop in value. Here we use the stress amplitude at which G' = G'' as the oscillatory yield stress, $\sigma_{y2}$ that represents energy dissipation from interparticle bond breaking under oscillatory shear. As can be seen in inset of Fig. \ref{fig:fig4} (b), oscillatory yield stress increased with increasing particle concentration with a power-law exponent of 2.4. Estimating yield strain (given by $\gamma_y$ = $\sigma_{y2}$/G') showed that it reduces with increasing particle concentration (not shown here). This indicates that the gel network is in the strong-link (or interfloc) regime as explained by Shih \textit{et al.}.\cite{shih1990scaling} In the strong-link regime the network breakup is dominated by bond breaking within a floc. Moreover, it is also possible that the flocs may interpenetrate depending on the preshear history\cite{dages2022interpenetration} and have a strong interfloc elastic constant. This observation is also supported by the value of d$_f$ calculated earlier that represents a strong-link regime for the AC particle gel. Interestingly both HB yield stress ($\sigma_{y1}$) and oscillatory yield stress ($\sigma_{y2}$) show similar power-law exponent when plotted against particle concentration as has been observed for carbon black gels.\cite{grenard2014timescales} These results indicate that despite the differences in particle morphology as well as surface chemistry, AC particles as well as carbon black particles exhibit similar structure-flow relationship highlighting the inherent nature of carbon-based anisotropic particles. 

\par Another important observation here is the absence of a second peak in G' and G'' beyond yield stress as has been reported in some of the colloidal gels.\cite{koumakis2011two,shukla2015two} In such systems the first peak is associated with interparticle bond breaking that generally occurs at lower stresses or strains and breaks the gel network dispersing the particle aggregates and making the gel behave like a ``cluster fluid''. An additional peak is observed at intermediate stresses or strains that are associated with restructuring of aggregates resulting in a highly heterogeneous gel with denser clusters.\cite{laurati2011nonlinear,koumakis2011two,moghimi2017colloidal} In our system we only observe a single yielding event implying breaking of aggregates from interparticle bond rupture.

\subsection{\label{sec:3.3} Surfactant impact on slurry elasticity}

Previous work has shown the addition of a nonionic surfactant can reduce viscosity and increase electrical conductivity in semi-solid electrodes containing carbon black particles.\cite{madec2015surfactant} We aimed to explore these effects for slurry electrodes based on AC particles used in aqueous flow batteries. This particular formulation question is especially challenging due to the irregular shape, hierarchical pore size distribution, and large surface area of AC particles. We measured the effect of surfactant addition on AC slurry linear and nonlinear viscoelastic response by systemtically adjusting the surfactant concentration for a fixed particle concentration of 14\% w/w. The ratio of nonionic surfactant to the concentration of particles (or $\alpha$) varied from 0 - 0.7.
\par
\begin{figure*}[h]
\includegraphics[width=0.9\linewidth]{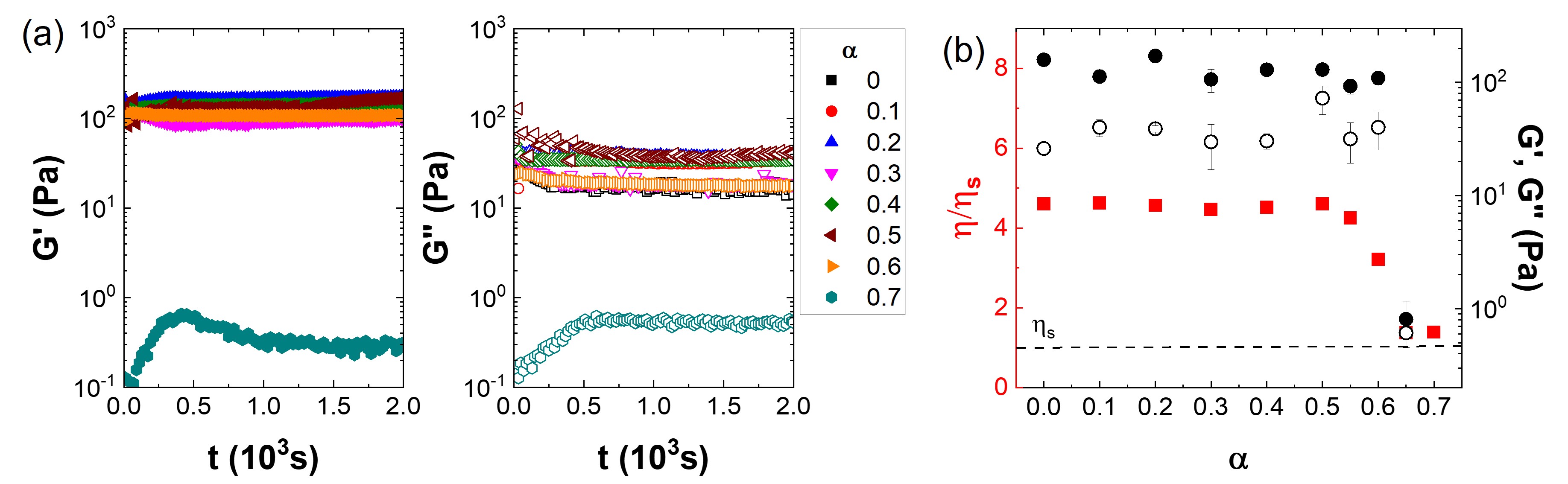}
\caption{\label{fig:fig5} (a) 14\% w/w sample ageing (or time-evolution of storage (G') and loss modulus (G'')) after high shear rejuvenation at different $\alpha$ (b) Comparison of suspension relative viscosity ($\eta/\eta_s$) and storage modulus G' (filled symbol, measured at $\omega$ = 10 rad.s$^{-1}$, $\gamma$ = 0.2\%) as a function of $\alpha$. The sample response transitions from a solid-like to liquid-like at $\alpha \ge$ 0.65.}
\end{figure*}

 Changing interparticle interaction through surfactant addition can influence particle aggregation after flow cessation. Hence, we first captured the aging of slurry after cessation of high shear rejuvenation. As seen in Fig. \ref{fig:fig5} (a), we observed that for $\alpha \le$ 0.6 the slurry behaves like a gel with elastic modulus being greater than viscous modulus. A similar behavior was observed during frequency sweep (Fig. S3). Interestingly, the values of elastic and viscous moduli as well as the kinetics of aging are similar to the slurry without added surfactant. However, at $\alpha \ge$ 0.65 the values of G' drop significantly and the slurry no longer behaves like a gel but like a liquid with G' < G''. This drop in slurry elasticity is abrupt unlike reported for carbon black suspensions containing surfactants where the drop was more gradual.\cite{khalkhal2018evaluating} In Fig. \ref{fig:fig5} (c) we plotted the values of shear moduli (measured at $\omega$ = 10 rad.s$^{-1}$ and $\gamma$ = 0.2\%) at the end of aging (t$_w$ = 2000 s) and relative viscosity during rejuvenation ($\dot{\gamma}$ = 1000 s$^{-1}$) and see that at $\alpha \ge$ 0.65, there is a clear gel-like to fluid-like transition. Note here that the solvent viscosity $\eta{_s}$ is the viscosity of the filtered supernatant measured after particle sedimentation using a centrifuge. Suppression of interparticle attraction greatly reduced slurry viscosity where at $\alpha$ = 0.7, the slurry viscosity is only slightly above the solvent viscosity.

\subsection{\label{sec:3.4} Surfactant impact on slurry yielding}

\begin{figure*}[ht]
\includegraphics[width=0.6\linewidth]{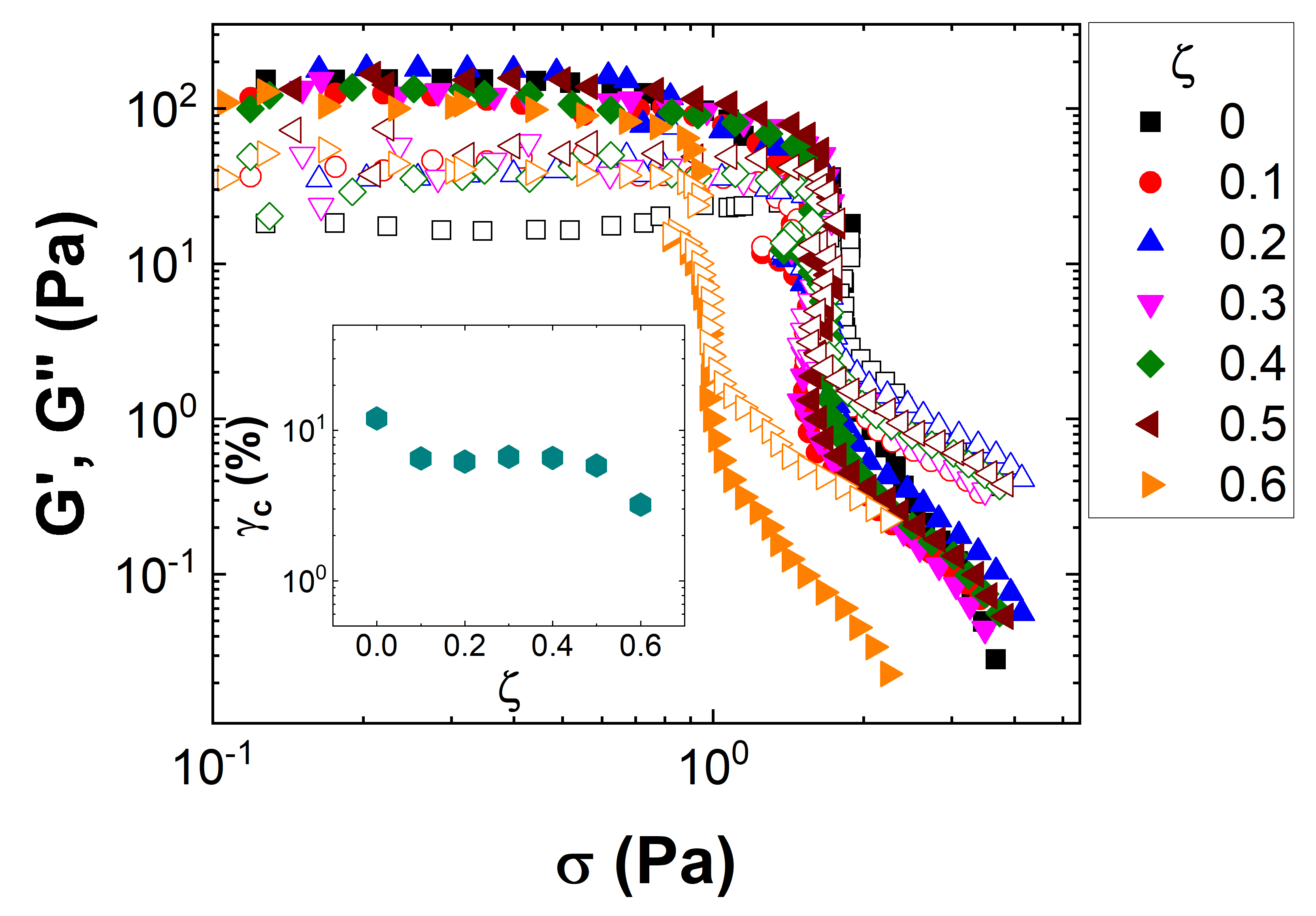}
\caption{\label{fig:fig6} Elastic (G') and viscous modulus (G'') measured during DSS test (at $\omega$ = 10 rad.s$^{-1}$) for a sample of 14\% w/w AC slurry with increasing surfactant concentration plotted as a function of measured stress $\sigma$. Inset: Cross-over strain $\gamma_c$ ($\gamma$ when G' = G'') plotted as a function of $\alpha$ = c$_{surf}$/c$_{CB}$.}
\end{figure*}

In the case of colloidal gels, altering interparticle attraction potential leads to changes in network elasticity and is reflected in their yielding (or solid-like to liquid-like transition) behavior.\cite{laurati2011nonlinear} Understanding interparticle potential is more complicated when particles have shape anisotropy, size polydispersity and rough surface morphology.\cite{fejer2011self,monti2019effect} We look at the yielding behavior by plotting elastic (G') and viscous modulus (G'') as a function of measured stress $\sigma$ with increasing strain amplitude $\gamma$ (Fig. \ref{fig:fig6}) for slurry samples with increasing surfactant concentration. Our results show that for $\alpha \le$ 0.5 there is no significant change in yield stress $\sigma_y$ (stress when G' = G'') indicating similar magnitude of gel network elasticity independent of surfactant concentration. There is a slight drop in slurry yield stress at $\alpha$ = 0.6 with similar values of G' and G'' as other samples. Increasing $\alpha \ge$ 0.7 results in complete fluidization of slurry samples. Similar to yield stress, the cross over strain $\gamma_c$ ($\gamma$ when G' = G'') shows a slight drop at $\alpha$ = 0.6. It is known that yielding in colloidal gel networks is a result of local particle rearrangement within a cluster with increasing strain/stress amplitude.\cite{koumakis2011two,laurati2011nonlinear} Small values of $\gamma_c$ indicate that the slurry gel network consists of large particle flocs which break under very small deformation. However, of crucial importance here is the fact that significant increase in c$_{surf}$ did not alter slurry yield stress/strain. 

\subsection{\label{sec:3.5} Surfactant effect on slurry steady shear flow}
\begin{figure*}[ht]
\includegraphics[width=0.7\linewidth]{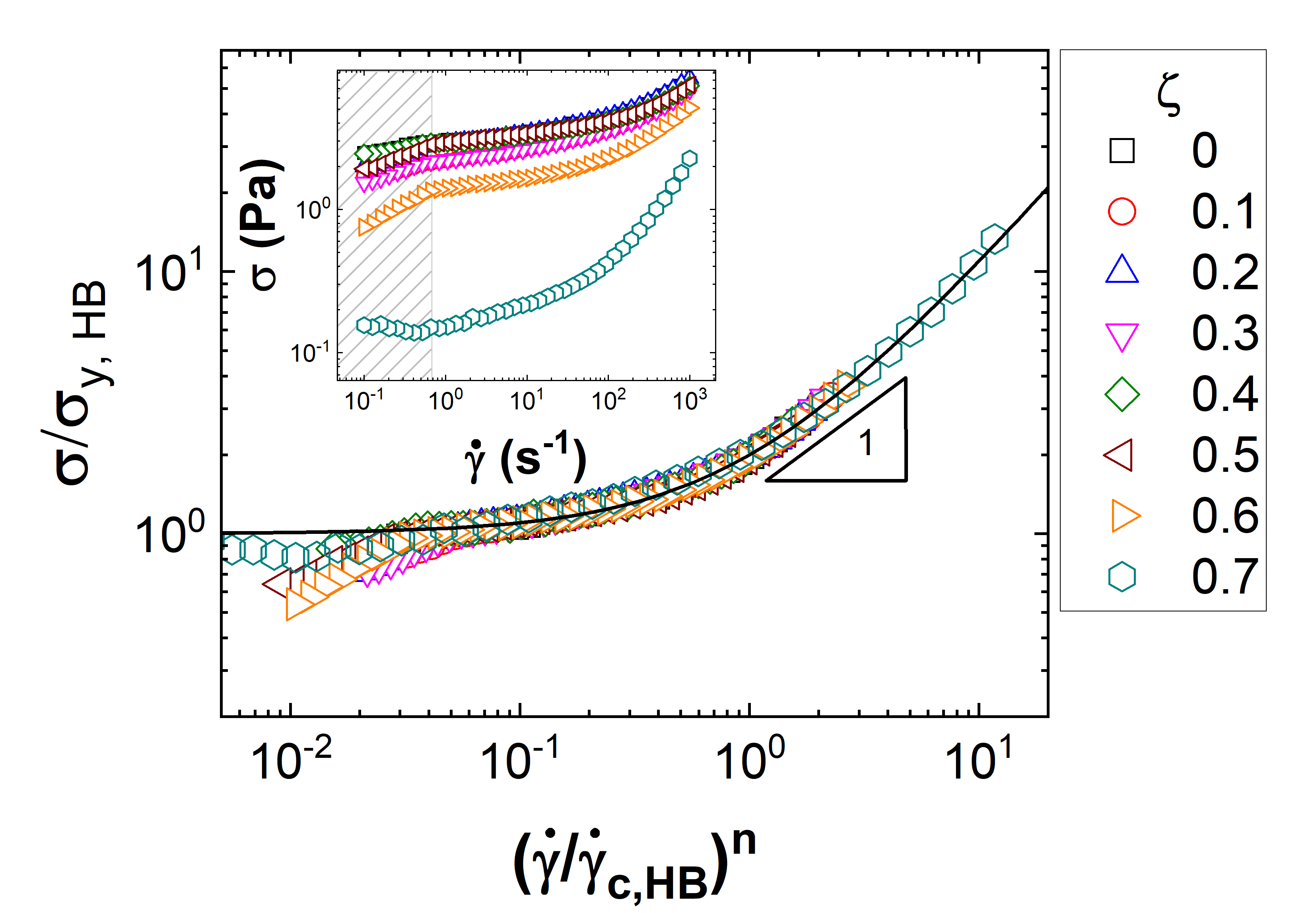}
\caption{\label{fig:fig7} Flow curve for 14\% (w/w) slurry samples with different surfactant concentration plotted on a dimensionless plot using the Herschel-Bulkley fitting parameters as shown in Table \ref{tab:table2}. Inset shows flow curves without normalization.}
\end{figure*}

Next, we measured the flow curve to understand the effect of surfactant addition on the 14\% (w/w) slurry that transitions from a fluidized state at high shear rate to a solid-like state at very low shear rates as shown in Fig. \ref{fig:fig2}. As reported earlier, addition of surfactant can reduce slurry viscosity and improve carbon particle dispersion.\cite{madec2015surfactant} However, our results (Fig. \ref{fig:fig7}) show that similar to linear rheology, there is no significant change in measured shear stress as a function of applied shear rate for $\alpha \le$ 0.6 whereas at $\alpha$ = 0.7 there is a sharp drop in measured stress and viscosity in the range of applied shear rates (Fig. \ref{fig:fig7} inset). There is drop in shear stress at the lowest applied shear rates which we attribute to rheopexy or antithixotropy\cite{ovarlez2013rheopexy,wang2022new} as explained in Section III (b). Flow curves of slurry samples with different surfactant concentrations collapse on to a single master curve (Fig. \ref{fig:fig7}) defined for self-similar structure flow curves shown by Hipp \textit{et al.} for carbon black gels\cite{hipp2021direct} and defined here using Eq. \ref{Eq:Eq1}. Hipp \textit{et al.} defined $\sigma/\sigma_y$ as inverse Bingham number or Bi$^{-1}$ whereby at Bi$^{-1}$ > 1 the slurry is in the strong flow regime with reversible thixotropy and at Bi$^{-1}$ < 1 it is in the weak flow regime where antithixotropy dominates. In the strong flow regime the particle clusters are broken down into smaller aggregates and the slurry is fluidised whereas in the weak flow regime, densification of interpenetrating particle clusters takes place that reduces their hydrodynamic volume causing a sharp drop in slurry viscosity as was also shown in Fig. \ref{fig:fig2} (b). At $\alpha$ = 0.7, the slurry viscosity drops significantly, however, the master curve shows that the mechanism behind particle restructuring at different shear rates is similar for slurry samples with increasing surfactant concentrations.
\par At this point it is important to note that increased surfactant concentration will cause steady loss of free particle surface area necessary to form interparticle bonds. However, instead of an incremental drop in gel elasticity, apparent yield stress and viscosity, we see an abrupt drop in these parameters above a critical surfactant concentration of $\alpha$ = 0.7. These data suggest a potentially different mechanism from that of surfactant adsorption onto the free surface of the AC particles. We further explored the nature of surfactant adsorption.

\begin{table}[ht]
\caption{\label{tab:table2}Parameters used for Herschel-Bulkley model fits to the suspension flow curve }
\begin{ruledtabular}
\begin{tabular}{ccdd}
$\alpha$ & $\sigma_y$ (Pa) & \mbox{$\dot{\gamma_c}$} & \mbox{n} \\
\hline
0 & 2.9 $\pm$ 0.02 & 371 & 0.5 \\
0.1 & 2.1 $\pm$ 0.02 & 194 & 0.5 \\
0.2 & 2.8 $\pm$ 0.02 & 291 & 0.5 \\
0.3 & 2.1 $\pm$ 0.02 & 215 & 0.5 \\
0.4 & 2.8 $\pm$ 0.02& 423 & 0.5 \\
0.5 & 3 $\pm$ 0.02 & 471 & 0.55 \\
0.6 & 1.4 $\pm$ 0.01 & 196 & 0.6 \\
0.7 & 0.17 $\pm$ 0.005 & 65 & 0.9 \\
\end{tabular}
\end{ruledtabular}
\end{table}
\subsection{\label{sec:3.6}Surfactant adsorption on AC particles}

Surfactant adsorption was measured as described in a previous section with UV-vis spectroscopy. The surfactant used here, Triton X-100, show sharp peaks at wavelengths of 230 nm and 275 nm with a shoulder around 280 nm.  \cite{Nedaei2021} The free surfactant concentration in the supernatant of carbon black slurry samples was measured by tracking the magnitude of these peaks (for absorbance plot, please refer to Fig. S4, ESI). Supernatant of samples with $\alpha$ < 0.7 only showed large broad peaks from 200-250 nm and from 280-350 nm range which are peaks that represent iron complexes that form when dispersed in aqueous solutions.\cite{loures_advanced_2013} However, for samples with $\alpha$ $\ge$ 0.7, the two peaks at 230 nm and 275 nm with a shoulder start to appear, clearly indicating the presence of surfactants in the supernatant. This surfactant concentration regime agrees with our observations from the rheological study, whereby we concluded that at $\alpha$ $\ge$ 0.7 our particle free surface is saturated by surfactant monomers resulting in suppression of interparticle attraction and the resulting gelation. Additionally, our previous study showed that AC particle gels exhibit a catastrophic collapse under gravitational stress at $\alpha$ > 0.75 further supporting our current observation. \cite{LEE2023712}

\par We calculated the surface area coverage of surfactants to carbon black particles at the critical ratio of $\alpha$ = 0.7. Estimating that the head group of the surfactant molecule adsorbing to the AC surface occupies 68 \AA$^{2}$/molecule \cite{parra2020effect} and the surface area of YP-50F to be 1600 m$^{2}$/g, the calculated AC surface area coverage by surfactant monomers is approximately 30\%. This shows that approximately 30 \% coverage of the surface area of the AC particles by surfactant molecules is sufficient to induce interparticle repulsion that results in fluidization of the slurry. It needs to be noted here that the surface area measured using BET methods are always larger for highly porous materials such as AC particles.\cite{gonzalez2000determination} However, due to the large size of the surfactant monomer when compared to the nitrogen molecule used in BET measurement, the entire BET surface area of AC particles is not accessible to the surfactant.\cite{gonzalez2000determination} In addition to the AC particle surface area, the surface chemistry also has an important role in influencing surfactant adsorption. Oxygen containing surface groups on AC particles deter adsorption of hydrophobic molecules and show greater affinity towards water molecules.\cite{wu2001adsorption,pendleton2002activated} This can strongly affect regions where surfactant monomers adsorb on the particle surface creating areas of varying surfactant monomer concentration. However, in such a scenario one would expect more and more surfactant rich areas on particle surface with increasing surfactant concentration in the solution and gradual decrease in interparticle attraction which is not the case in the present study.

\par Another possibility is that at $\alpha$ < 0.7, surfactant monomers adsorb into the pores of the AC particle surface whereas the external surface remains free of surfactants. Studies on ionic and nonionic surfactant adsorption on AC particles have shown that the surface porosity influences surfactant adsorption more than surface chemistry.\cite{gonzalez2001influence,gonzalez2004ionic} Firstly, surfactant adsorption occurs through monomer adsorption and not through micellar adsorption. Secondly, at the lowest equilibrium surfactant concentrations, the monomers adsorb onto the highly energetic sites such as the micro pores. As surfactant concentration is increased, the monomer adsorption occurs at lower energetic sites such as the meso and macro-pores.\cite{gonzalez2001influence,gonzalez2002removal,krivova2013adsorption} In such a scenario, interparticle bonds can form between AC particles through their external surface until surfactant concentration is high enough to saturate the micro-pores and start adsorbing on to the external surface. It cannot be ruled out that in the present study the adsorbent-adsorbate interaction is influenced both by surface chemistry as well as porosity.

\subsection{\label{sec:3.7}Slurry conductivity}

\begin{figure*}[ht]
\includegraphics[width=0.7\linewidth]{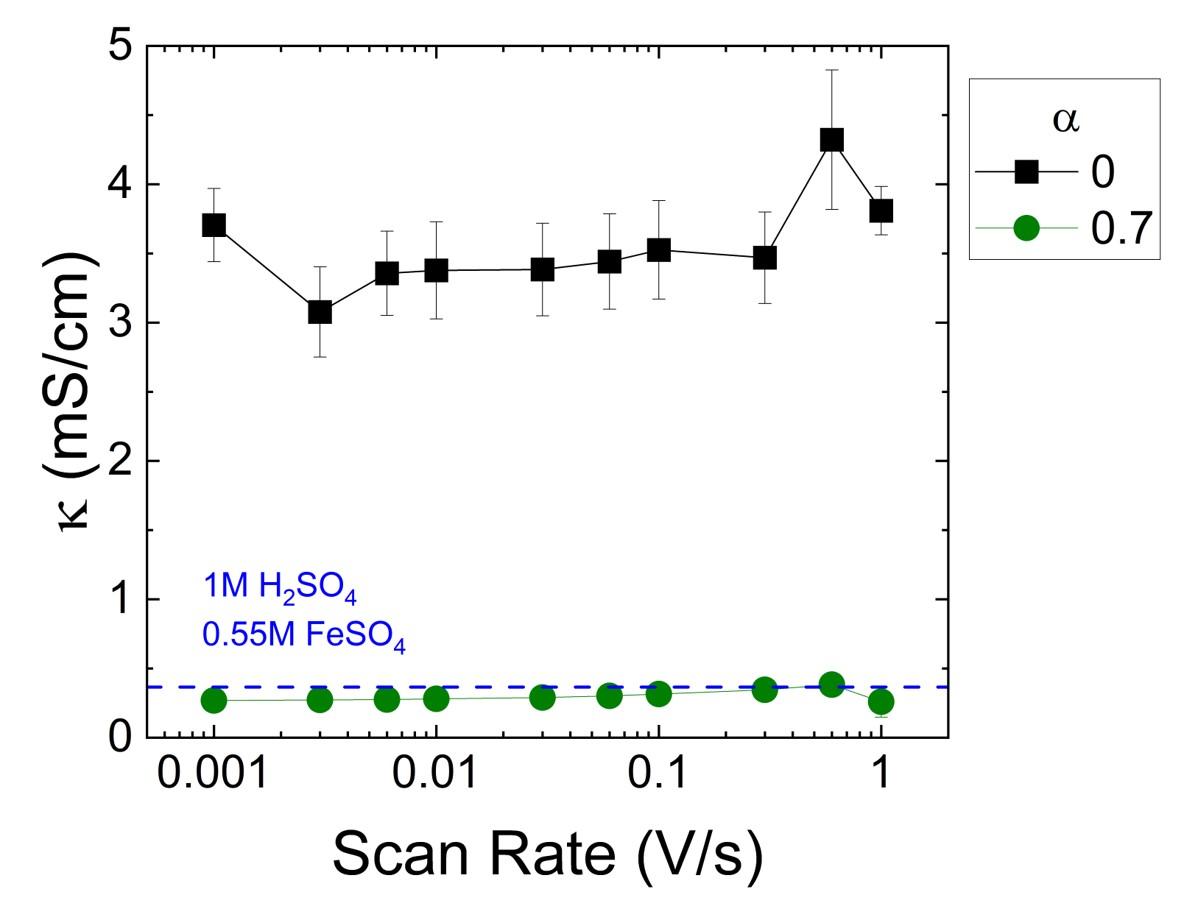}
\caption{\label{fig:fig8} Conductivity of slurry flowing through the cell at 170 mL/min with no surfactant and at surfactant saturation at varying scan rates.}
\end{figure*}

\par The electrical performance of the slurries is central to our work given the application of these materials. The aim of our work is to not only understand how formulation choices impacts the stability and rheological response of slurries, but also how the electrical properties may be impacted.
The conductivity of the slurry measured at different scan rates at $\alpha$ = 0 and 0.7 are presented in figure \ref{fig:fig8}. The dotted blue line represent the conductivity measured with no AC particles and just the solvent (1 M H$_2$SO$_4$ and 0.55 M Fe$_2$SO$_4$). Without surfactants ($\alpha$ = 0), the conductivity of the slurry ranges between 3-4 mS/cm. However, at surfactant saturation ($\alpha$ = 0.7), the conductivity drops to around 0.25-0.35 mS/cm which is close to the conductivity of the electrolyte solvent at 0.35 mS/cm, suggesting that the AC particle's contribution in the overall conductivity is completely removed. While further work will be needed to fully understand this change in the electrical behavior of these particles with addition of surfactants, it is interesting to see that the contribution of AC particles to the overall conductivity also diminishes at the same surfactant ratio at which we observe the loss of gel-like behavior. Clearly, surface saturation of AC particles by surfactant monomers not only disrupts the mechanical percolation, but also the electrical percolation.

\section{\label{sec:5}Conclusions}
\par In conclusion, our rheological investigation reveals that AC particle slurry used in flow battery applications are yield-stress materials behaving like the out-of-equilibrium soft systems known as gels. The power-law scaling of the elastic modulus with particle concentration reveals a fractal-like structure for the AC particle gel when in a quiescent state. A nonionic surfactant (Triton X-100) was added to the AC particle slurry with an intention to modify the slurry viscoelasticity and flowability. At $\alpha$ (=c$_{surf}$/c$_{AC}$ < 0.65), there is no change in slurry viscosity or elasticity, however, at $\alpha \ge$ 0.65, the slurry abruptly transitions from a gel to a fluid. We attribute this behavior to the selective adsorption of surfactant molecules into the pores of the AC particles at $\alpha$ < 0.65 rather than their external surface. Such a phenomenon does not alter interparticle attraction. Instead of increasing AC particle dispersion and improving electrical conductivity, addition of the nonionic surfactant results in complete loss of conductivity also at $\alpha \ge$ 0.7 and renders the slurry unsuitable for electrochemical applications.

\par Our study highlights the profound and abrupt impact of nonionic surfactant on the rheological response of AC  slurry electrodes. These results suggest formulation choice, along with the unique particle surface morphology, could have unintended effects on slurry viscoelasticity and electrical properties. Ongoing work in our lab aims to elucidate additional details on how electrical and mechanical percolation are related, and how both features are impacted by slurry formulation. We hope this and future studies will help researchers better understand the underlying structural mechanisms affecting flow battery performance.

\nocite{*}
\bibliography{aipsamp}

\end{document}


\title{\textbf{Surfactant-Driven Dynamic Changes in Rheology of Activated Carbon Slurry Electrodes}\\ Supplementary Information} 

\author{Mohan Das}
\altaffiliation[Also at ]{Faculty of Physics, University of Vienna, Boltzmanngasse 5, A-1090, Vienna, Austria}
\author{KangJin Lee}%

\author{Christopher L. Wirth}
 \email{wirth@case.edu.}
\affiliation{
Department of Chemical and Biomolecular Engineering, Case Western Reserve University, Cleveland, Ohio 44106.
}%

\date{\today}

\maketitle 


%
%
\section{Nonlinear rheology}
We performed shear startup and creep measurements on a 14\% w/w AC slurry to capture the steady state response of shear stress and shear rate respectively. During shear startup, the measured stress shows a decay at long time indicating antithixotropic dynamics of the system. Formation of interpenetrating particle flocs at very low shear rates results in their densification and subsequent reduction in their hydrodynamic size.
\par During creep, the gel-like network formed by the AC particles is able to withstand the applied stresses until when the network yields at higher stresses and results in a rapid fluidization of the slurry.

\begin{figure*}[ht]
\includegraphics[width=0.5\linewidth]{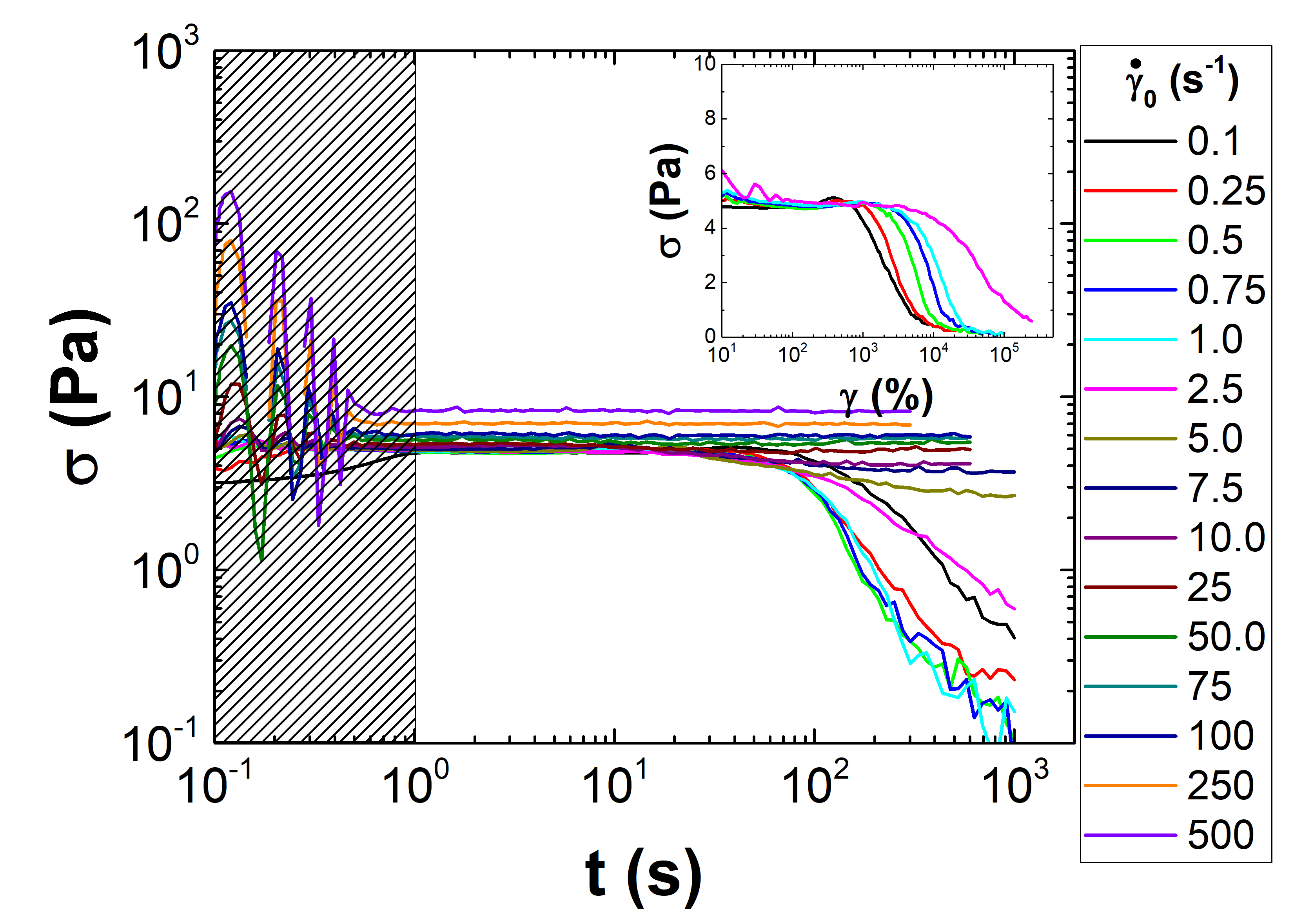}

\caption{\label{fig:figS1} Shear stress $\sigma$ recorded at constant applied shear rate $\dot{\gamma_0}$ during shear startup measurement on AC slurry at C = 14 \% w/w. Shaded region represents short time instrument error. Inset: Same data but for the lowest applied shear rates.}
\end{figure*}

\begin{figure*}[ht]
\includegraphics[width=0.5\linewidth]{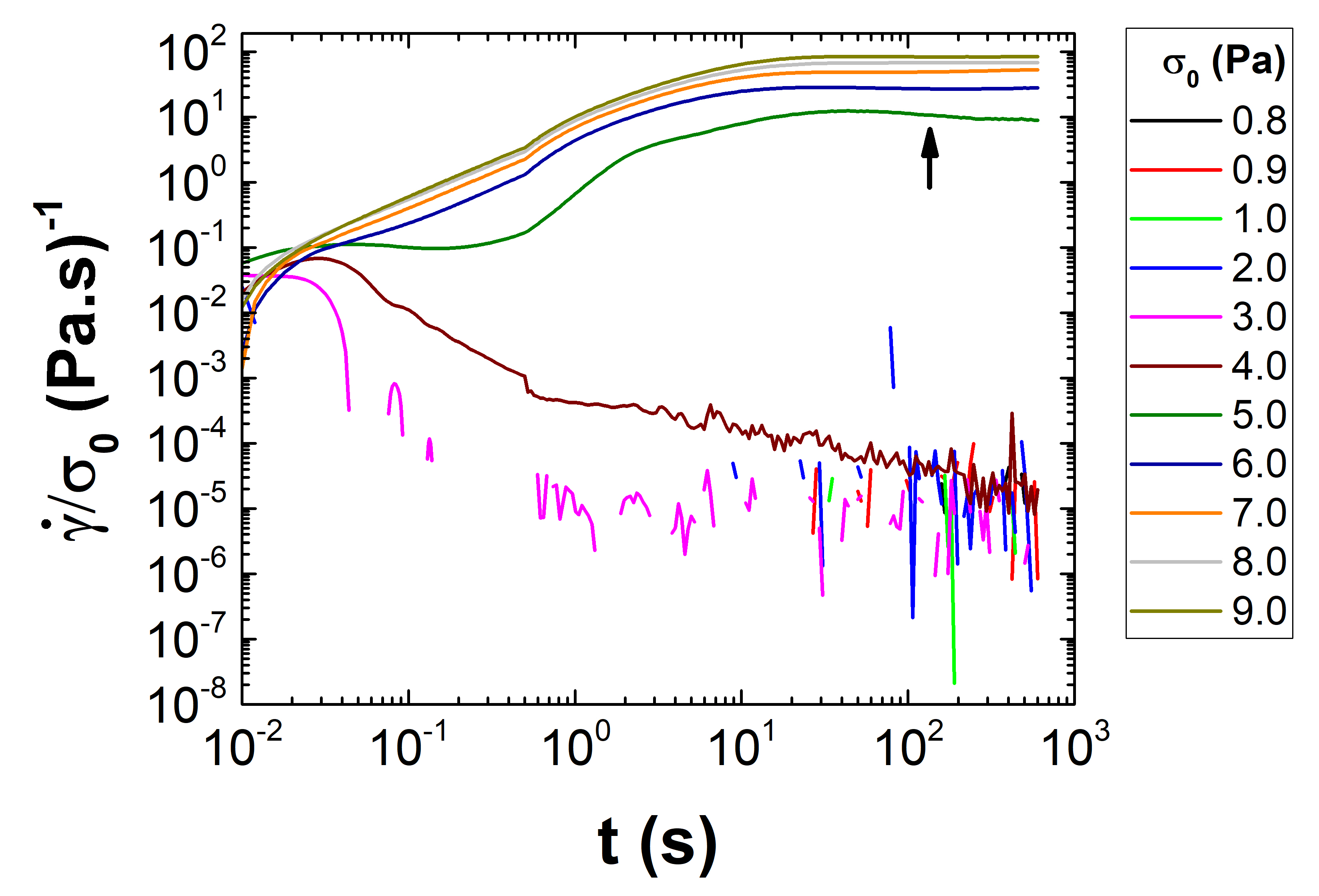}

\caption{\label{fig:figS2} Normalized shear rate $\dot{\gamma}$/$\sigma_0$ recorded at constant applied shear stress $\sigma_0$ during creep measurement on AC slurry at C = 14 \% w/w. Arrow represents the regime of abrupt fluidization of the AC slurry sample.}
\end{figure*}

\section{Linear rheology}
Linear rheological measurements such as dynamic frequency sweep can capture particle dynamics at different length scales. Typically in colloidal gels, the dynamics is solid-like in the measurable range of applied frequencies and elastic modulus exhibits a weak dependence on frequency. We observe a similar behavior in our samples with added surfactants until $\alpha$ < 0.7. This indicates the inability of surfactant monomers to alter interparticle interaction at different length scales at lower concentrations. 
\begin{figure*}[ht]
\includegraphics[width=0.5\linewidth]{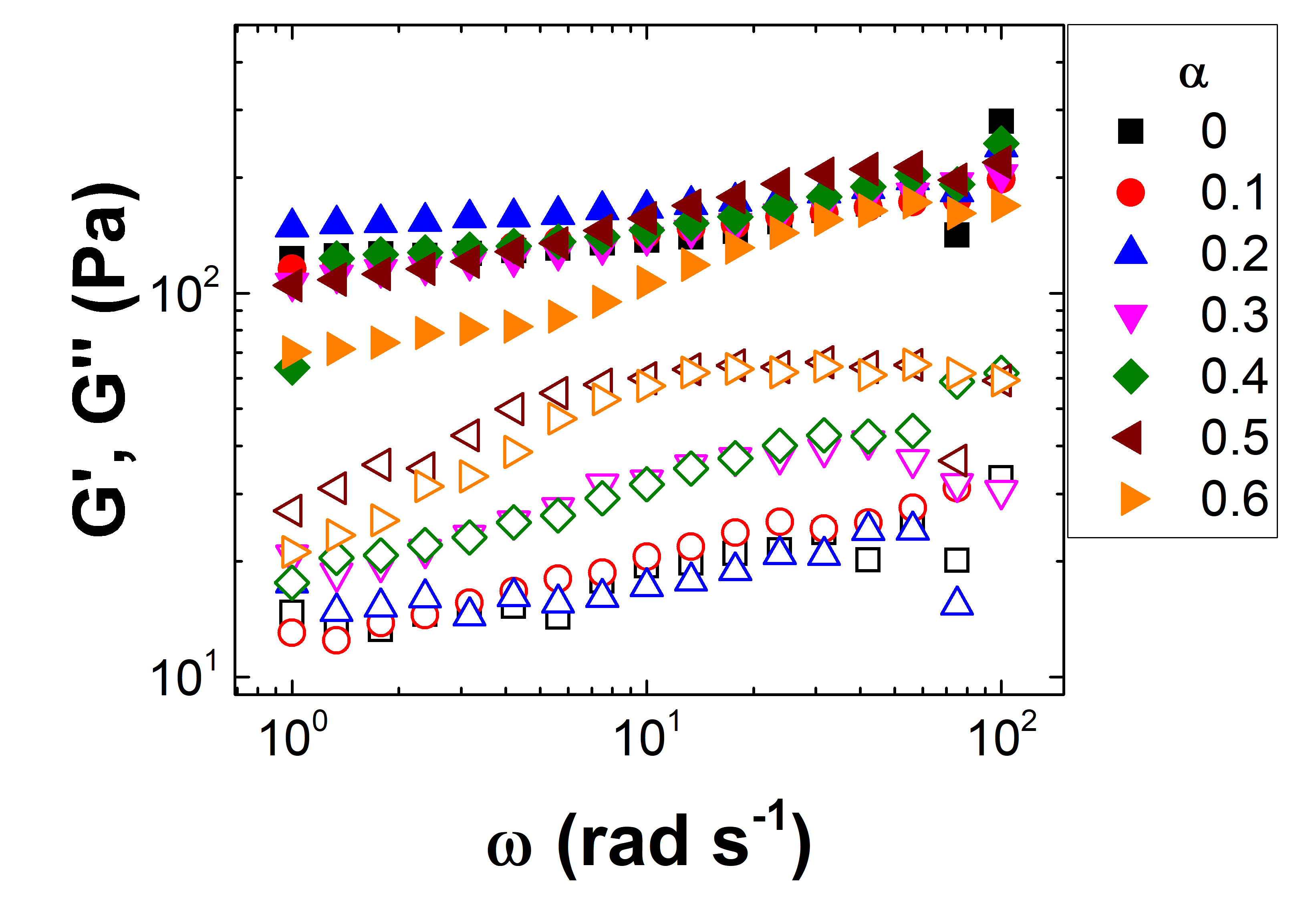}
\caption{\label{fig:figS3} Normalized shear rate $\dot{\gamma}$/$\sigma_0$ recorded at constant applied shear stress $\sigma_0$ during creep measurement on AC slurry at C = 14 \% w/w. Arrow represents the regime of abrupt fluidization of the AC slurry sample.}
\end{figure*}

\section{UV-Vis absorption of the supernatant}
In effort to characterize the adsorption of Triton X-100 to carbon black particles dispersed in 1M H$_2$SO$_4$ and 0.55M FeSO$_4$, supernatant of the mixed samples were measured via UV-vis spectroscopy. As described in the manuscript, the characteristic peak of Triton X-100 at 230nm and 275nm wavelength start to show up for samples at $\alpha$ $\ge$ 0.7 while samples at $\alpha$ < 0.7 show the iron complex peaks ranging from 200-250nm and 280-350nm. \cite{Nedaei2021,loures_advanced_2013} This suggests that the saturation of surfactants adsorbed to the carbon black particle surface happens around $\alpha$ = 0.7 which falls in the same range as our previous study where the suspension has the same formulation but just without the FeSO$_4$. \cite{LEE2023712}

\begin{figure*}[ht]
\includegraphics[width=0.5\linewidth]{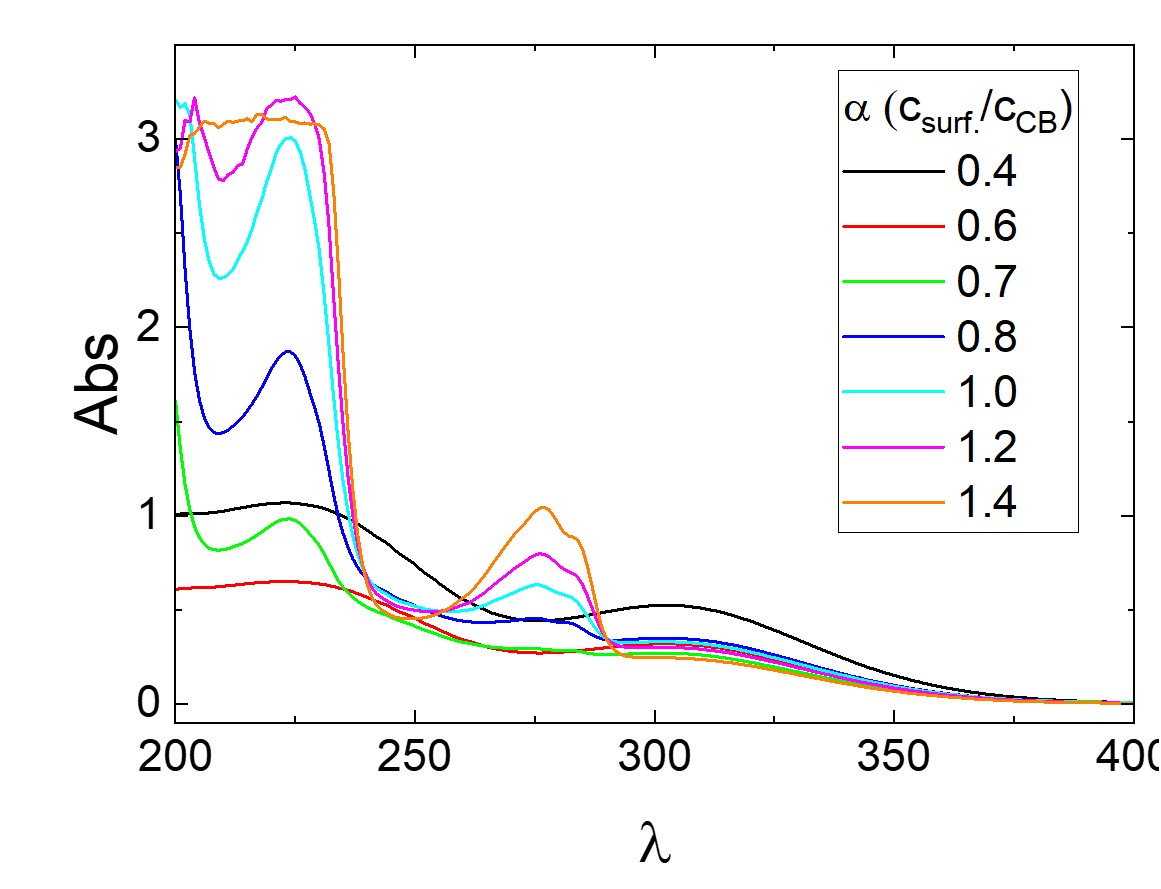}

\caption{\label{fig:figS4} Absorbance measured from the supernatant collected from samples at $\alpha$ from 0.4 to 1.4.}
\end{figure*}

%



\bibliography{aipsampSI}